\documentclass[aps,prl,twocolumn]{revtex4}

\usepackage{latexsym}
\usepackage{amsmath}
\usepackage{amsfonts}
\usepackage{amssymb}
\usepackage{amsbsy}
\usepackage{bm}
\usepackage{color}
\usepackage[dvips]{graphicx}

\graphicspath{{./figures/}}

\begin{document}
\begin{titlepage}
\title {Flow-Induced Helical Coiling of Semiflexible Polymers
in Structured Microchannels}
\author{Raghunath Chelakkot$^\dagger$$^\ddagger$, Roland G. Winkler$^\dagger$, Gerhard Gompper$^\dagger$}
\affiliation{
$^\dagger$Theoretical Soft Matter and Biophysics, Institute of
Complex Systems and Institute for Advanced Simulation,
Forschungszentrum J\"ulich, 52425 J\"ulich, Germany \\
$^\ddagger$Physics Department, Brandeis University, Waltham, MA
02453, USA }

\begin{abstract}
The conformations of semiflexible (bio)polymers are studied in
flow through geometrically structured microchannels. Using
mesoscale hydrodynamics simulations, we show that the polymer
undergoes a rod-to-helix transition as it moves from the narrow
high-velocity region into the wide low-velocity region of the
channel. The transient helix formation is the result of a
non-equilibrium and non-stationary buckling transition of the
semiflexible polymer, which is subjected to a compressive force
originating from the fluid-velocity variation in the channel. The
helix properties depend on the diameter ratio of the channel, the
polymer bending rigidity, and the flow strength.
\end{abstract}

\maketitle
\end{titlepage}

Buckling  is a common phenomena of slender bodies, like long
filaments and thin sheets, under an external load. Examples range
from microscopic to macroscopic length scales, and include actin
filaments, vaulting poles, virus capsids, and tectonic plates.
Specifically, slender rods or semiflexible (bio)polymers exhibit a
buckling instability under compression, when the load exceeds a
critical value \cite{kroy:96,eman:07,kier:10}. In the simplest
situation of Euler buckling of a rod, the symmetry ($O_2$ symmetry
along the rod axis) is brocken by buckling perpendicular to its
axis in an arbitrary direction \cite{land:86,golu:00}. Since
buckling is of such fundamental importance, it has received
persistent attention over centuries.

Typically, buckling transitions are considered under equilibrium
conditions. For macromolecular and biological filaments with
typical length scales of nano- to micrometers, which we are
interested in, thermal fluctuations broaden the buckling
transition and generate a smooth crossover from the unbuckled to
the buckled state \cite{kroy:96,eman:07,kier:10}.  However, very
little attention has been payed to conformational instabilities
far from equilibrium \cite{golu:00,geng:09}, which exhibit
qualitatively new features compared to equilibrium ones. Here, the
transport of semiflexible polymers in microchannels and
capillaries presents a new opportunity to study the
non-equilibrium behavior of such filaments
\cite{chel:10,stei:12,kant:12}. Vice versa, a detailed
understanding of the dynamical process involved in such a
transport is of paramount importance in many applications. This
applies, in particular, to many biologically-relevant polymers,
such as DNA, actin filaments, and microtubules, which are
semiflexible; an example is DNA sorting in microchannels
\cite{chou99,jo:07,levy10}.

\begin{figure}
\begin{center}
\begin{tabular}{ll}
\quad(a)&\qquad(b)\\
\includegraphics*[width=0.48\columnwidth,clip]{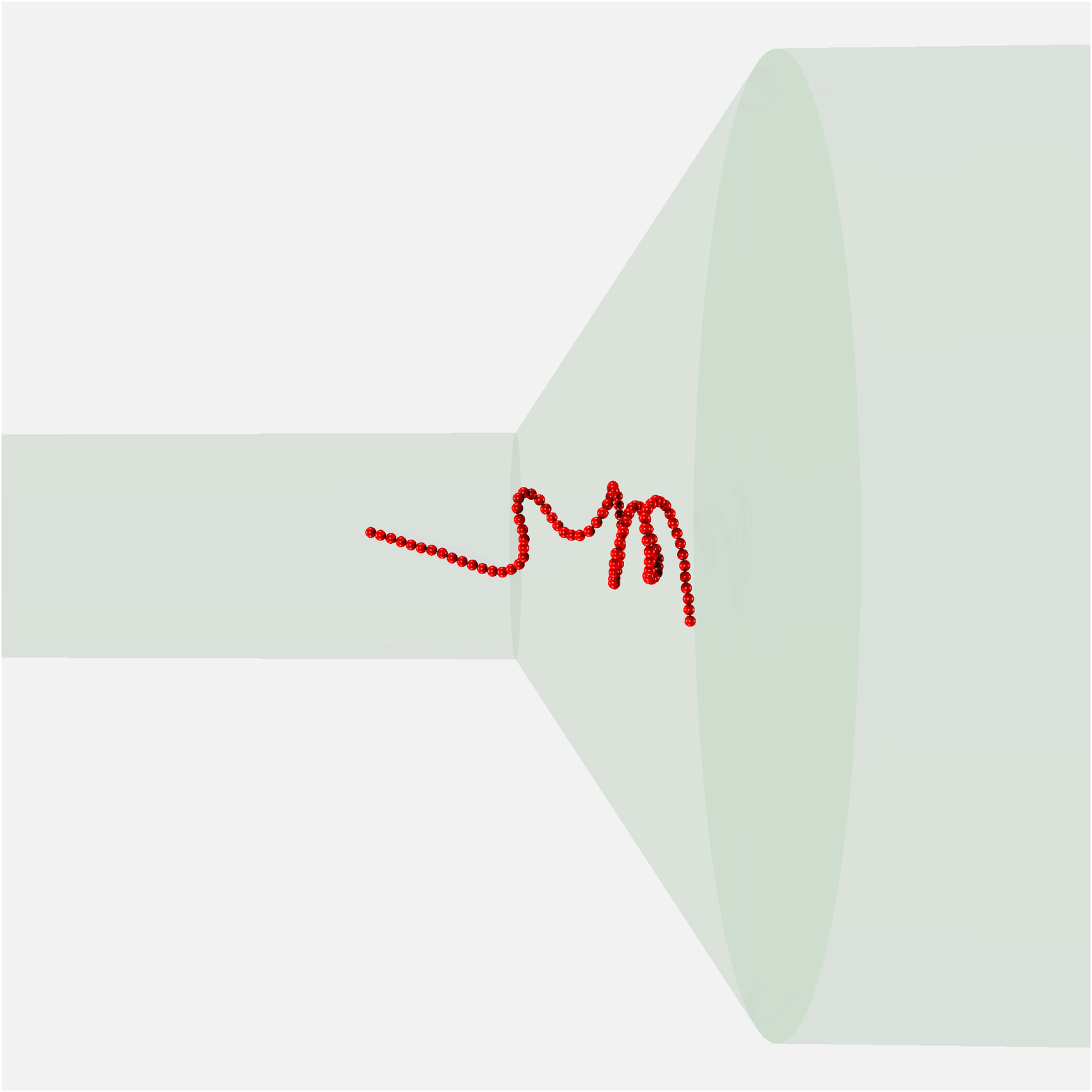}&
\includegraphics*[width=0.48\columnwidth,clip]{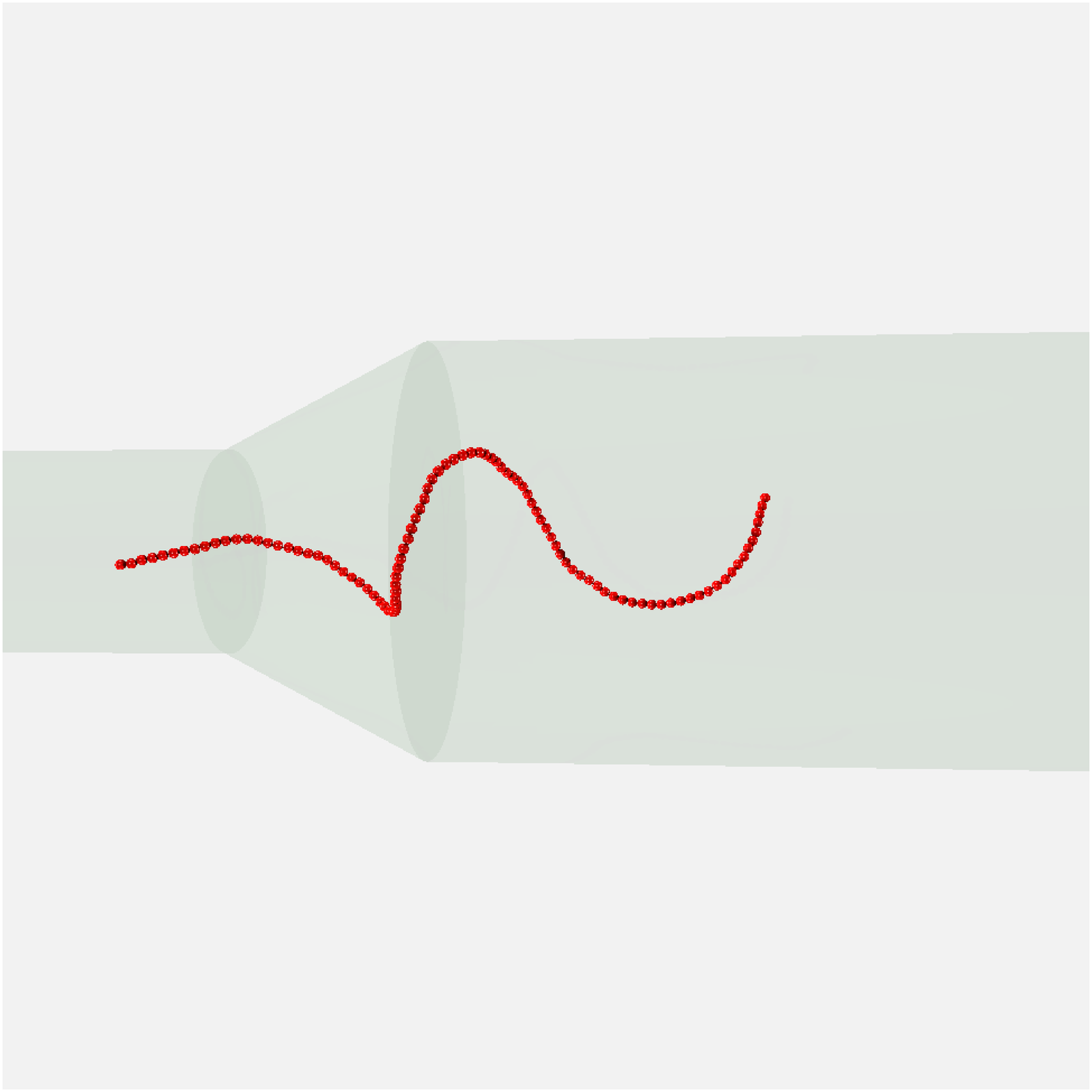}
\end{tabular}
\caption{(color online) Polymer conformations in channels with the diameter ratios
(a) $D_w/D_n=2$ and (b) $D_w/D_n=4$ for the flow strength $Pe\simeq550$ and
$L/L_p =1$. See also movies S1 and S2 in
the Supplemental Material \cite{movie}.
} \label{ex_channel}
\end{center}
\end{figure}

Non-equilibrium instabilities can appear under various conditions.
Here, we investigate flow fields of spatially varying flow
strength. Such a situation is easily realized in flows through
spatially-structured microchannels, {\em e.g.}, in a region where
the channel width changes from narrow to wide (see
Fig.~\ref{ex_channel}). By mesoscale hydrodynamic simulations of a
semiflexible polymer in such a microchannel, we observe a buckling
of the polymer as it enters the wider channel section. Buckling is
often the first step in the formation of more complex structures
\cite{mari:00,snir:05}. Indeed, we observe that buckling is
followed by a flow-induced helical coiling of the polymer.

Helices are a preferred shape in nature. They are frequently
adopted in biological systems such as proteins, they have been
shown to be close to optimal packing under the restriction of
their own volume and finite length \cite{mari:00}, they can arise
for entropic reasons in the presence of depletant molecules
\cite{snir:05}, and bacteria swim by rotating helical filaments
\cite{berg:73}. Helices are also formed by falling ropes and
falling fluid filaments \cite{maha:96,maha:98}, and transiently in
homopolymer collapse from an extended conformation \cite{sabe08},
or are induced by electrostatic interactions in polyelectrolytes
\cite{xu:09}. However, as we will explain, the helix formation
process in microchannel flow is different from previously studied
cases, because it occurs at small Reynolds numbers, in the absence
of attractive interactions, and is mainly due to solvent friction.

We apply a hybrid simulation method, in which  molecular dynamics
simulations for semiflexible polymers are combined with a
mesoscale simulation method, the multiparticle collision dynamics
(MPC) approach, for the solvent \cite{kapr:08,gomp:09}. In the MPC
algorithm, the fluid is described by a set of $N$ point particles
of mass $m$, with velocities determined by a stochastic process.
The particle dynamics evolves in two steps. In the streaming step,
the solvent particles move ballistically for a time interval $h$.
In the collision step, particles are sorted into the cells of a
cubic lattice of lattice constant $a$ and their relative
velocities, with respect to the center-of-mass velocity of each
cell, are rotated around a randomly oriented axis by an angle
$\alpha$.  For every cell, mass and momentum are conserved, which
leads to the build up of hydrodynamic interactions (HI) between
the fluid particles; at the same time thermal fluctuations are
taken into account \cite{gomp:09,kapr:08,male:99}.

The fluid is confined in a cylindrical channel of diameter $D_w$,
with periodic constrictions of smaller diameter $D_n$ (see
Fig.~\ref{ex_channel}). In order to create a smooth laminar flow,
there is a segment between the narrow and wide parts, in which the
diameter interpolates linearly between $D_n$ and $D_w$. The length
of the constriction is larger than the polymer contour length $L$.
No-slip boundary conditions are imposed on the channel walls by
the bounce-back rule and virtual wall particles \cite{gomp:09}.
Flow is induced by a gravitational force ($mg$) acting on every
fluid particle in the direction of the channel axis, which yields
parabolic flow profiles; in the parts connecting the narrow and
wide segments, there is also a radial flow component. The
difference in velocity between the narrow and wide parts of the
channel is $|v_n| -|v_w| = |v_n|(1-D_n^2/D_w^2)=\Delta v$, where
$v_w$ and $v_n$ denote the velocities in the wide and narrow
segments, respectively.

The polymer is represented by a bead-spring model, in which $N_m$
monomers, each of mass $M$, are linearly connected by harmonic
springs of equilibrium length $b$ \cite{huan:10_1}. Excluded
volume interactions are taken into account by a purely repulsive
(truncated and shifted) Lennard-Jones potential
\cite{huan:10_1,gomp:09}. Additionally, a three-body potential is
applied to capture bending stiffness, with the bending rigidity
$\kappa$ \cite{chel:10}.

The polymer-fluid interaction is taken into account by including
the monomers in the collision step \cite{gomp:09}. We employ the
MPC parameters $h=0.1 \sqrt{ma^2/k_BT}$, $\alpha = 130^{\circ}$,
the mean number of particles per collision cell $\langle N \rangle
=10$, and the fluid mass density $\varrho = \langle N \rangle m
/a^3$. For the polymer, we set $M= m \langle N \rangle$, $b=a$,
and the Lennard-Jones parameters $\sigma=b$ and $\epsilon/k_BT=1$.
If not otherwise indicated, the polymer length is $L/b =N_m-1=99$.
To maintain a constant temperature, a local velocity scaling
algorithm is applied, which yields Maxwell-Boltzmann distributed
velocities. We characterize the strength of the flow by the Peclet
number $Pe=\dot \gamma \tau$, where $\dot\gamma = g \varrho D_n/(4
\eta)$ is the shear rate at the cylinder wall and $\tau$ the
longest relaxation time of a semiflexible polymer \cite{harn:96}.
Based on a semiflexible polymer model, we find the relaxation time
$\tau/\sqrt{ma^2/k_BT} \approx 1.4 \times 10^{6} L/L_p$ for the
above parameters, where $L_p =\kappa/(k_BT)$ is the persistence
length of the polymer.

Figure~\ref{ex_channel} shows snapshots of polymer conformations
when the polymer leaves the constriction and enters the wider part
of the channel. Inside the constriction, the polymer conformations
are rodlike with small fluctuations. When the polymer is ejected
from the constriction, it enters a region where the  fluid
velocity decreases. Hence, it is subjected to a compressive force
as a result of the difference in fluid velocity, and the polymer
undergoes conformational changes. With increasing flow strength,
we find a gradual change in transient polymer conformations from
(bend) rods to helical shapes. This helical conformation is a
novel feature for a semiflexible polymer, and occurs by
spontaneous symmetry breaking. Simulations for various channel
diameters $D_w$, but fixed constriction diameter $D_n$, show that
the helix diameter strongly depends on $D_w$ and decreases with
increasing channel diameter ratio $D_w/D_n$, see
Fig.~\ref{ex_channel}. We attribute this behavior to the larger
compressive force due to the larger difference $\Delta v$ in fluid
velocities for larger $D_w/D_n$. It is evident from the snapshots,
in particular in Fig.~\ref{ex_channel}b, that the polymers assume
helical shapes without any direct interaction with the channel
walls. Hence, any qualitative influence of confinement can be
unequivocally ruled out as the dominating mechanism.

\begin{figure}
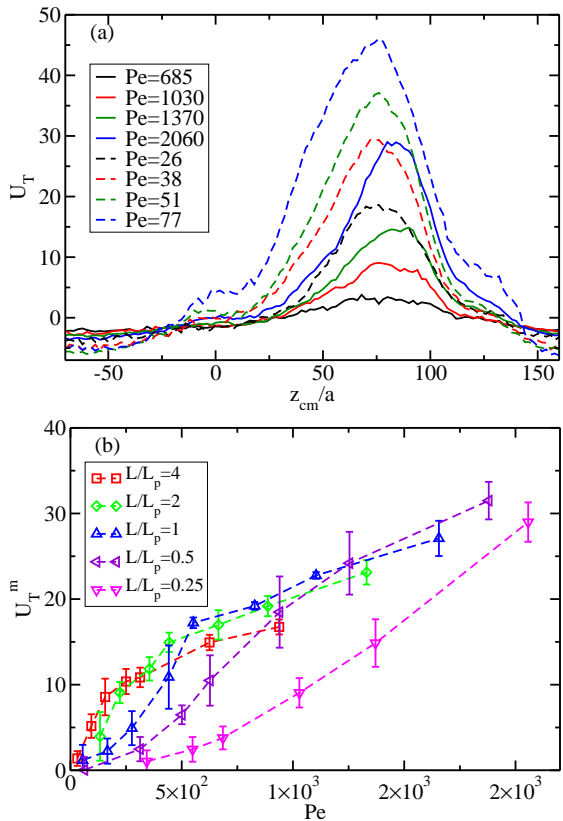

\begin{center}
\begin{tabular}{l}
\includegraphics*[width=0.85\columnwidth,clip]{tor_energy.eps}\\
\includegraphics*[width=0.85\columnwidth,clip]{tor_energy_max.eps}
\end{tabular}
\caption{(color online) (a) Order parameter $U_T$ as function of the axial
position of the polymer center of mass for systems with (solid lines)
and without (dashed lines) hydrodynamic interactions. The flow
strengths are $Pe = 685,1030,1370,2060$ (HI)
and $Pe=26,38,51,77$ (no HI),  the
ratio $L/L_p=1/4$, and  $D_w/D_n=2$.
(b) Peak values $U_T^{m}$ of the order parameter $U_T$
for the persistence lengths $L/L_P= 4.0 \ (\square), \ 2.0 \ (\Diamond), \
1.0 \ (\triangle), \ 0.5 \ (\lhd),$ and $0.25 \ (\triangledown)$.
The results are for the MPC fluid with $D_w/D_n=2$. \label{fig:U_t}}
\end{center}
\end{figure}

In order to understand the underlying mechanisms and driving
forces, we first study the influence of hydrodynamic interactions.
Simulations with turned-off hydrodynamic correlations using the
Brownian MPC approach~\cite{gomp:09}, where each monomer
independently performs stochastic collisions with uncorrelated
fluid particles subject to a parabolic flow profile, provide the
same qualitative behavior. Hence, the rod-to-helix transition is
not caused by hydrodynamic interactions, although substantial
quantitative differences in helical properties are observed.

To quantify the helical conformations, we introduce the torsional
order parameter $U_T = \sum_{i=1}^{N_m-1}  \cos\gamma_i$, where
the angle $\gamma_i$ is defined by the three subsequent bond
vectors ${\bm R}_{i-1}$, ${\bm R}_{i}$, and ${\bm R}_{i+1}$, with
\begin{equation}
 \cos\gamma_i = \frac{({\bm R}_{i-1}\times {\bm R}_{i}) \cdot
 ({\bm R}_{i}\times {\bm R}_{i+1})}
{|{\bm R}_{i-1}\times {\bm R}_{i}||{\bm R}_{i}\times {\bm R}_{i+1}|} .
\end{equation}
$U_T$ is displayed in Fig.~\ref{fig:U_t}(a) for the persistence
length $L_p=4L$. $U_T$ increases as the polymer forms a helix when
it is ejected from the constriction, and decreases again as the
polymer relaxes back to its original straight configuration.
Simulations without hydrodynamic interactions display a very
similar behavior except that the values of $U_T$ are significantly
larger; this is due to the lack of fluid being dragged along by
the polymer segments.

Figure~\ref{fig:U_t}(b) presents the maximum values $U_T^m$ of the
average order parameter $\left\langle U_T(z)\right\rangle$ for
various stiffnesses in the range $1/4 \le L_p/L \le 4$. The
parameter of more flexible polymers increases faster than that of
stiffer ones. All peak values seem to saturate at high Peclet
numbers, where the saturation value increases with increasing
stiffness. Within a certain range of Peclet numbers, a universal
dependence of $U_T^m$ appears independent of stiffness.

\begin{figure}[t]
\begin{center}
\includegraphics*[width=0.95\columnwidth,clip]{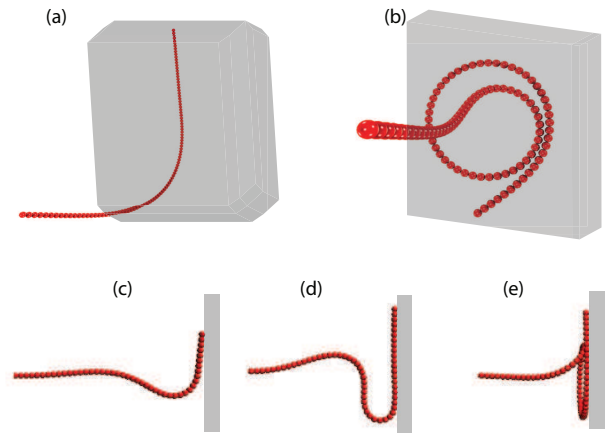}
\caption{(color online) Typical configurations of buckled semiflexible polymers with
$L_p/L = 1.0$. The rigid substrate is indicated by the gray area.
The polymer length is $N_m=100$ and a the force is $F/(k_BT/a)=100$.
The viscous drag coefficient is
(a) $\Gamma/\sqrt{m k_B T/b^2} =1.0$ and (b) $50.0$.
The sequence of images (c) - (d) illustrating the helical coiling of the
semiflexible polymer during the initial phase. See also movies S3 and S4 in
the Supplemental Material \cite{movie}}
\label{buckling_wall}
\end{center}
\end{figure}

In order to elucidate the factors which cause helical coiling of a
polymer under a non-homogeneous compressive forces, we consider a
simpler system, where a polymer in a viscous fluid is driven by a
rigid wall (penetrable to the fluid), which moves with the
constant velocity $v_w$. Thermal fluctuations and hydrodynamics
interactions are omitted for simplicity. The planar wall is
parallel to the $xy$-plane and perpendicular to the polymer, which
is oriented initially along the $z$-direction with one end in
contact with the wall. The polymer dynamics is governed by
\begin{equation}
M \ddot{\bm r}_i=-\Gamma \dot{\bm r}_i + {\bm F}_i^w \Theta(z_w-z_i)
                         + {\bm F}_i^l +{\bm F}_i^b,
\label{eq:1}
\end{equation}
where ${\bm r}_i$ is the position of monomer $i$,  $z_w$ the
position of the wall, $\Theta(z)$ the Heaviside step function,
${\bm F}_i^l$ and ${\bm F}_i^b$ are the bond and bending forces,
respectively, and $\Gamma$ is the monomer friction coefficient.
${\bm F}_i^w = F {\bm e}_z$, where ${\bm e}_z$ is the unit vector
along the $z$-axis, describes the force of the wall on monomer
$i$, when it is in contact with the wall. Equation~\ref{eq:1} is
solved numerically for a polymer with $L_p/L = 1$.

\begin{figure}[t]
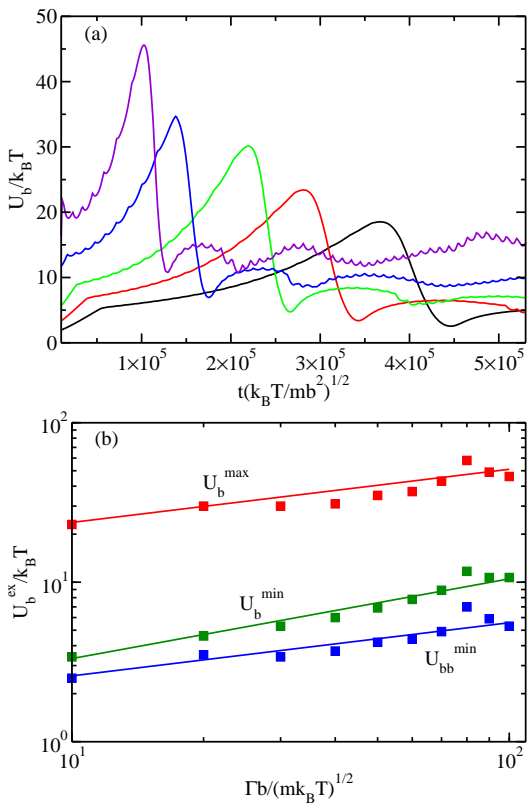

\begin{center}
\includegraphics*[width=0.8\columnwidth,clip]{bend_energy_time.eps}
\includegraphics*[width=0.8\columnwidth,clip]{bend_energy_ex.eps}
\caption{(color online) (a) Time dependent bending energy $U_b$
of a semiflexible polymer with $L/L_p=1.0$
pushed forward by a rigid wall moving with velocity
$v_w=0.1 \sqrt{k_BT/m}$.
Curves from left to right correspond to decreasing drag coefficients
$\Gamma/\sqrt{m k_B T/b^2} =100, 50, 20, 10, 5$.
(b) Dependence of the maximum bending energy $U_b^{max}$ (top), the minimum energy
after the drop $U_b^{min}$ (middle), and the minimum bulk energy $U_{bb}^{min}$
(bottom) on friction. The slopes of the straight lines are $1/3$ (top, bottom) and
$1/2$ (middle).} \label{ub}
\end{center}
\end{figure}

For small $\Gamma$, less than a threshold $\Gamma_c^{(1)}$,
frictional dissipation is weak and the polymer moves with the wall
without any significant conformational changes. For fixed ${\bm
v}_w$, the behavior changes qualitatively when $\Gamma$ exceeds
$\Gamma_c^{(1)}$, because the compressive force due to polymer
friction now leads to a buckling transition, similar to classical
Euler buckling. As the wall moves further, the part of polymer in
contact with the wall aligns with the wall and moves along the
wall, whereas the part further away from the wall moves in the
same direction, but a smaller velocity (cf.
Fig.~\ref{buckling_wall}a). Ultimately, the entire polymer is
stretched out parallel to the wall. With further increasing
$\Gamma$, beyond a threshold $\Gamma_c^{(2)}$, we observe a
distinctly different behavior, as the polymer now exhibits a
transition to a coiled state (cf. Fig.~\ref{buckling_wall}b). As
displayed in Fig.~\ref{ub}, the bending energy $U_b$ of the
polymer exhibits two different regimes as function of time for
such systems. Initially, $U_b$ increases by local Euler-like
buckling of the polymer over a length scale much smaller than its
length (cf. Figs.~\ref{buckling_wall}(c), \ref{ub}(a)). When the
bending energy exceeds a certain value, the polymer undergoes a
conformational change to a coiled state near the wall
(Fig.~\ref{buckling_wall}(d),(e)), and $U_b$ drops significantly.
For later times, the polymer continues coiling until the entire
chain is coiled up at the wall.

The dependence of the peak value of the bending energy on the wall
velocity and the friction coefficient $\Gamma$ can be understood
from some scaling arguments. For $\Gamma < \Gamma_c^{(1)}$, the
motion of the polymer through the viscous medium generates a
inhomogeneous (negative) tension $\sigma(s) = \Gamma v_w (L-s)$,
there $s$ is the arc length measured from the polymer end at the
wall. For a homogeneous tension $\sigma_0$
buckling theory predicts an instability with a fastest growing
mode with wave vector $q \sim \sqrt{\sigma_0/\kappa}$, hence
buckling occurs when $q > 2\pi/L$. If we approximate the tension
by its average $\sigma_0 = \Gamma v_w L/2$, we obtain
$\Gamma_c^{(1)} \sim \kappa v_w^{-1} L^{-3}$. \\
For friction coefficients $\Gamma > \Gamma_c^{(2)}$, initial
buckling occurs at the near-wall end of the polymer
(Fig.~\ref{buckling_wall}(1)). The bending energy of the buckled
part of contour length $L_c$ has the local curvature $1/R(s)$ and,
hence, the bending energy is $U_b \sim \kappa \int_0^{L_c} R^{-2}
ds$. Since there is only local bending of the polymer over a
length scale $L_c \approx R$, we find $U_b \sim \kappa/R$ by
replacing $R(s)$ by a characteristic value $R$. The resulting
bending force $F_b = - \kappa/R^2$ is balanced by the frictional
force over the same length scale, {\em i.e.}, $F_f= \Gamma v_{w}
R$, which implies a curvature radius
\begin{equation} \label{radius}
R \sim \left[\kappa/(\Gamma v_w) \right]^{1/3}
\end{equation}
and a bending energy
\begin{equation} \label{b_energy}
U_b \sim \kappa^{2/3} (\Gamma v_w)^{1/3}.
\end{equation}

During the transition from local buckling to circular coiling, the
curvature of the buckled part decreases, which explains the drop
in bending energy (Fig.~\ref{buckling_wall}(2),(3)). At later
times, during the stationary coiling process, the dynamics is
determined by two contributions: the bending energy of a part
$L_c$  not in contact with the wall $U_{bb}$(bulk part) and a part
of circular shape at the wall. For the bulk part of the bending
energy, the same argument applies as in the derivation of
Eqs.~(\ref{radius}), (\ref{b_energy}). Hence, we expect to find
the same scaling relation for bending in the bulk.

Figure \ref{ub}(b) shows the scaling behavior with respect to
friction of the maximum bending energy $U_b^{max}$ of
Fig.~\ref{ub}(a). This energy nicely follows the scaling
prediction (\ref{b_energy}).  To characterize the the
quasi-stationary behavior of the bending energy for long times, we
determine the minimum energy $U_{bb}^{min}$ of the bulk part of
the bending energy after the steep drop of $U_b$. The numerical
simulations show that the bulk bending energy dominates over the
wall part for small friction coefficients. At the larger $\Gamma$
the two contributions become comparable. As shown in
Fig.~\ref{ub}(b), $U_{bb}^{min}$ also follows the scaling relation
(\ref{b_energy}), whereas the total minimum energy exhibits the
stronger dependence $\Gamma^{1/2}$.

The buckling and helical coiling of a polymer under flow in a
structured microchannel and a polymer pushed forward by a wall in
a viscous medium is governed by very similar mechanisms. In
contrast to the moving-wall model, the polymer in the structured
microchannel does not exhibit circular collapsed conformations.
Indeed, the compressive force due to the fluid velocity difference
is less restrictive, because the coiled part of the polymer can
relax more easily. The observed behavior differs from classical
Euler buckling in several respects.  First, the buckling force
appears due to a non-equilibrium and non-stationary dynamical
process. Second, buckling is initiated locally. This is a
consequence of the frictional force, which increases with the
length of the considered polymer part. Third, a transition to
helical coiling occurs when the local tilt angle exceeds a certain
value.

Thus, helical coiling is expected to be a generic phenomenon of
polymer motion in non-homogeneous viscous environments. Actin
filaments \cite{stei:12} in structures microchannels \cite{nogu10}
seem to be an optimal system to observe the helical-coiling
transition experimentally.


\end{document}